\documentclass[11pt]{article}

\usepackage{amsmath}
\usepackage{amsfonts}
\usepackage{indentfirst}
\usepackage{graphicx}
\usepackage[paper=letterpaper,lmargin=3.0cm,rmargin=3.0cm,tmargin=3.0cm,bmargin=4.0cm]{geometry}
\usepackage{sectsty}

\sectionfont{\large}
\subsectionfont{\normalsize}

\def\me#1{\begin{equation}\begin{split}#1\end{split}\end{equation}}
\def\men#1{\begin{equation*}\begin{split}#1\end{split}\end{equation*}}
\def\meg#1{\begin{subequations}\begin{align}#1\end{align}\end{subequations}}
\def\se#1{\begin{equation}#1\end{equation}}
\def\sen#1{\begin{equation*}#1\end{equation*}}
\def\pmx#1{\begin{pmatrix}#1\end{pmatrix}}

\def\m{\textrm{ mod }}
\def\mod#1{\ (\textrm{mod }#1)}
\def\i{\textrm{i}}
\def\fontMCG{\mathsf}
\def\A{\fontMCG{A}}
\def\B{\fontMCG{B}}
\def\S{\fontMCG{S}}
\def\E{\fontMCG{E}}
\def\U{\fontMCG{U}}
\def\W{\fontMCG{W}}
\def\C{\fontMCG{C}}

\begin{document}

\title{\LARGE Mapping class group and $\mathrm{U}(1)$ Chern-Simons theory\\ on closed orientable surfaces}
\author{Si Chen\footnote{E-mail: sichen@phas.ubc.ca}\\
\\
\it Department of Physics and Astronomy,\\
\it University of British Columbia,\\
\it 6224 Agricultural Road,
\it Vancouver, BC V6T 1Z1, Canada}
\maketitle

\begin{abstract}
$\mathrm{U}(1)$ Chern-Simons theory is quantized canonically on manifolds of the form $M=\mathbb{R}\times\Sigma$, where $\Sigma$ is a closed orientable surface. In particular, we investigate the role of mapping class group of $\Sigma$ in the process of quantization. We show that, by requiring the quantum states to form representation of the holonomy group and the large gauge transformation group, both of which are deformed by quantum effect, the mapping class group can be consistently represented, provided the Chern-Simons parameter $k$ satisfies an interesting quantization condition. The representations of all the discrete groups are unique, up to an arbitrary sub-representation of the mapping class group. Also, we find a $k\leftrightarrow1/k$ duality of the representations.
\end{abstract}

\section{Introduction and Results}

Quantization of Chern-Simons gauge field theory has garnered much interest in the last twenty years. There are several different ways that the theory can successfully be quantized, and they relate diverse areas in mathematics and physics. The path-integral approach~\cite{Witten:1988hf,Moore:1989yh} shed light on the relation between Chern-Simons theory, knot invariants and conformal field theory. Geometric quantization~\cite{Axelrod:1989xt} gave a general three-dimensional quantization. Canonical quantization can be performed using either a real polarization~\cite{Elitzur:1989nr}, or a complex polarization and coherent states~\cite{Bos:1989kn,Gukov:2004id}, and a general theory has been developed using quantum groups~\cite{Alekseev:1994pa,Alekseev:1995,Meusburger:2003wk}.

Chern-Simons is a topological field theory which does not depend on details of the geometry of the space-time on which it is defined, rather only the topological features. In particular, since its action is an integral of 3-form, it is invariant under diffeomorphisms of the underlying space-time. Like gauge transformations, diffeomorphisms split into two types, small diffeomorphisms which can be continually deformed to the identity and large diffeomorphisms which cannot. The quotient group of all the diffeomorphisms by small diffeomorphisms is called mapping class group (MCG). In Chern-Simons theory, small diffeomorphisms are equivalent with small gauge transformations on shell, and can be fixed at the classical level. On the other hand, MCG, as well as the large gauge transformations (LGT) group, are discrete symmetries of the theory and there is no requirement that quantum states should be invariant. Instead, they should carry representations of these groups. It is generally an interesting question to ask what representations are allowed. In this Paper, we shall address this question in the context of $\mathrm{U}(1)$ Chern-Simons theory.

The example of Chern-Simons theory with a $\mathrm{U}(1)$ gauge group is particularly tractable in that quantization can be done very explicitly.  When the spacetime manifold is the product $M=\mathbb{R}\times\Sigma$, with $\mathbb{R}$ the time and $\Sigma$ a two-dimensional orientable surface, explicit wave-functions can be obtained~\cite{Bos:1989wa,Manoliu:1996fx} and shown to have an interesting relationship to rational conformal field theory.

In this paper, we shall revisit the quantization of $\mathrm{U}(1)$ Chern-Simons theory on closed orientable surfaces. Our goal is to examine role of the MCG of the surface in quantizing the theory. We find the following results. We shall demonstrate that the MCG is quantizable and we shall find its representation explicitly. If we seek quantization with a finite dimensional Hilbert space, we find that it is possible only when the parameter $k$ (in Eq.(\ref{ChernSimonsAction})) is a rational number.  This result is known from previous works \cite{Polychronakos:1990xq,Dunne:1998qy}. In these works and in Ref.~\cite{Bos:1989wa}, when $k=p/q$ with $p$ and $q$ coprime, it was stated that $p$ (or $k$ in~\cite{Bos:1989wa}) must be an even integer. Our results differ qualitatively from these quantizations and depend on the genus. When the genus is one ($\Sigma$ is the 2-torus), $k$ can be any rational number. For higher genus, one of $p$ or $q$ must be even. Moreover, by incorporating MCG, we find that for a given $k$, the representations of the holonomy group and LGT group become unique, and the representation for MCG is also unique, apart from an arbitrary unitary sub-representation which acts on the holonomy group and LGT group trivially.

Generally, at the classical level, when $\Sigma$ has genus $g$, the group of large gauge transformations is $\mathbb{Z}^{2g}$. We find that, commensurate with results of Ref.~\cite{Bos:1989wa}, this discrete group is realized undeformed at the quantum level only when $k$ is an integer. However, even in that case, we find that, augmenting the quantization with the requirement that the Hilbert space carries a unitary representation of the MCG, restricts the representation of the large gauge transformations to those where states are strictly invariant (theta angles associated with large gauge transforms vanish). Furthermore, we shall show that, when $k$ is rational but not integer-valued, the discrete group of large gauge transformations, which was abelian at the classical level, obtains a 2-cocycle and becomes a clock algebra~\cite{Polychronakos:1990xq,Dunne:1998qy}. We find an interesting $k\leftrightarrow 1/k$ duality of the representations of the homology group and the large gauge group which, with the restrictions on $k$ stated above, is compatible with our quantization of the MCG.

The reader might wonder why, in a topological field theory, where the action does not depend on the metric, the quantization of the MCG could be an issue at all. In order to do canonical quantization, we must choose a set canonical variables and, to quantize, we must further choose a polarization.  It is the latter which is not generally covariant. Then, covariance needs to be restored by quantizing the MCG.  As we shall show, the quantization of the MCG is non-trivial and, as we have discussed above, it can only be carried out with some restrictions on $k$ and even then it poses restrictions on certain parameters which arise naturally in the quantization of the theory. See also the discussions in~\cite{Peldan:1995rv,Matschull:1999he}.

This Paper is organized as follows. In Section 2 we will review the properties of classical Chern-Simons theory. Because MCGs for $\Sigma_1$, $\Sigma_2$ and $\Sigma_g, g\ge3$ have different explicit presentations, in Section 3, 4 and 5 we will quantize the phase space and then the MCG for those three cases respectively.

\section{General Formalism}

Chern-Simons theory with gauge group $\mathrm{U}(1)$ has the action
\se{
I_{\textrm{CS}}[A]=\frac{k}{4\pi}\int_{M}A\wedge dA,
\label{ChernSimonsAction}
}
which has gauge symmetry $A\rightarrow A+g^{-1}dg$, where $g$ is a $\mathrm{U}(1)$-valued function. We shall
consider the Hamiltonian approach on a 3-manifold $M=\mathbb{R}\times\Sigma$ where $\mathbb{R}$ is time and $\Sigma$ is a closed oriented 2-manifold. The 1-form field $A$ defined on $M$ can be decomposed as $A=A_t+A_\Sigma$,  with $A_t$ the temporal component and $A_\Sigma$ the components on $\Sigma^2$. The Chern-Simons action is decomposed as
\me{
I_{\textrm{CS}}[A]=\frac{k}{4\pi}\int_{M} A\wedge dA=&\frac{k}{4\pi}\int_{M}(A_t\wedge dA_t+A_\Sigma\wedge dA_t+A_t\wedge dA_\Sigma+A_\Sigma\wedge dA_\Sigma)\\
=&\frac{k}{4\pi}\int_{M}(A_\Sigma\wedge d_t A_\Sigma+2A_t\wedge d_\Sigma A_\Sigma)
\label{ChernSimons}
}
where $d_t$ and $d_\Sigma$ are the exterior differentiation operators on $\mathbb{R}$ and $\Sigma$ respectively. In the above action, $A_t$ acts as a Lagrange multiplier and enforces the constraint that the connection on $\Sigma$ is flat, $F_\Sigma=d_\Sigma A_\Sigma=0$. In the meantime, $A_t$ is integrated over. By an abuse of notation, we use $A$ to denote $A_\Sigma$ from now on.

By Hodge decomposition, any 1-form field $A$ can be written as
\sen{
A=dU+\bar dV+h
}
where $\bar d$ is the adjoint to $d$, $U$ is a 0-form, $V$ is a 2-form, $h$ is a harmonic 1-form, and $U,V,h$ are all $\mathrm{u}(1)$-valued. By the equation of motion $dA=0$, we get $V=0$. Because any $\mathrm{u}(1)$-valued function can be continuously deformed to zero, the term $dU$ can be eliminated by a small gauge transformation. There are still large gauge transformations to consider, which have gauge function with nonzero winding number around some non-trivial loop. To be explicit, for $\Sigma_g$, we can take a set of generators of the fundamental group $\bar\alpha_n,\bar\beta_n,n=1,\ldots,g$, such that $\#(\bar\alpha_n,\bar\alpha_m)=0,\#(\bar\beta_n,\bar\beta_m)=0,\#(\bar\alpha_n,\bar\beta_m)=\delta_{n-m}$, where $\#(,)$ is the algebraic intersection number between two loops. Then there is complete basis of harmonic 1-forms $\omega^{\alpha n}, \omega^{\beta n}$, such that $\oint_{\bar\alpha_n}\omega^{\alpha m}=\oint_{\bar\beta_n}\omega^{\beta m}=\delta_{n-m},\oint_{\bar\alpha_n}\omega^{\beta m}=\oint_{\bar\beta_n}\omega^{\alpha m}=0$, which implies $\int_{\Sigma^2}\omega^{\alpha n}\wedge\omega^{\beta m}=\delta_{n-m}$, and $\int_{\Sigma^2}\omega^{\alpha n}\wedge\omega^{\alpha m}=\int_{\Sigma^2}\omega^{\beta n}\wedge\omega^{\beta m}=0$. Since the field $A$ only has the harmonic part,
\se{
A=\sum_{n=1}^g\left(a_n\omega^{\alpha n}+b_n\omega^{\beta n}\right).
\label{ADecomp}
}
For $N_{\alpha n},N_{\beta n}\in\mathbb{Z}$, gauge transformations of gauge function
\sen{
g(x)=\exp\left[\sum_{n=1}^g\i2\pi\left(N_{\alpha n}\int_{x_0}^x\omega^{\alpha n}+N_{\beta n}\int_{x_0}^x\omega^{\beta n}\right)\right]
}
commute with the small gauge transformations, and are called large gauge transformations. Effectively they translate the variables $a_n,b_n$ by multiples of $\i2\pi$. They form the abelian group $\mathbb{Z}^{2g}$.

Substitute (\ref{ADecomp}) into (\ref{ChernSimons}), the action reduces to
\sen{
I_{\textrm{CS}}=\frac{k}{2\pi}\int dt\sum_{n=1}^ga_n\partial_tb_n.
\label{Defab}
}
According to the canonical quantization recipe, at this point, there is some freedom in choosing the canonical coordinate and momentum. We will use a real polarization here. Namely, $b_i$ and $\frac{k}{2\pi}a_i$ are taken as the canonical variables, with the commutation relation $[a_n,b_n]=\frac{-\i 2\pi}{k}$, and any other pairs commute.

We shall require the quantum states to transform under LGTs covariantly, and try to preserve classical properties of LGT as much as possible. Generators of LGTs can be written as $\rho_n=\exp(ka_n)$ and $\sigma_n=\exp(kb_n)$, and they label different Fourier modes of the wavefunction on the lattice. From the commutator of $a_n$ and $b_n$, we can find they satisfy the relation
\se{
\rho_n\sigma_n=\sigma_n\rho_n\exp(k^2[a_n,b_n])=\sigma_n\rho_n\exp\left(-\i2\pi k\right).
\label{ClockAlgebra}
}
This is called the clock algebra. When $k$ is not integer-valued, this is a deformed version of the classical commutation relation between $\rho_n$ and $\sigma_n$. It seems natural to interpret this deformation of algebra as a quantum effect. If one requires the quantum states to form a representation of the original undeformed classical algebra, then $k$ is quantized to be integer-valued. We will not make such requirement in this paper, and our results apply to the case of integer-valued $k$ straightforwardly.

Aside from carrying a representation of the above clock algebra, a quantum state also stores information that is invariant under LGT. The invariant subspace of the phase space is a $2g$-dimensional torus, parameterized by generators of the holonomy group $\alpha_n=\exp\left(\oint_{\bar\alpha_n}A\right)=\exp(a_n)$ and $\beta_n=\exp\left(\oint_{\bar\beta_n}A\right)=\exp(b_n)$. The non-trivial relation is another clock algebra
\se{
\alpha_n\beta_n=\beta_n\alpha_n\exp\left(-\frac{\i2\pi}{k}\right).
\label{DualClockAlgebra}
}
These two clock algebras (\ref{ClockAlgebra}) and (\ref{DualClockAlgebra}) can be regarded as being dual to each other, with duality transformation $k\leftrightarrow1/k$. Note that these two sets of operators commute. For example, $\alpha_n\sigma_n=\sigma_n\alpha_n\exp(-\i2\pi)$. Thus they realize the statement that on the classical level holonomies are invariant under LGTs.

Since both the LGT group and the holonomy group need to be represented in quantization, and there exists the above interesting duality transformation between them, we like to treat these two groups on an equal footing. This is another reason why $k$ is not restrained to be integer-valued. There is no obvious reason to insist that the holonomy group is undeformed, so by (\ref{DualClockAlgebra}), $1/k$ does not need to integer-valued. To exploit the duality, it is better to allow non-integer-valued $k$ as well.

To quantize the theory, we shall look for representations of the algebras (\ref{ClockAlgebra}) and (\ref{DualClockAlgebra}), as well as of the MCG with appropriate induced action on (\ref{ClockAlgebra}) and (\ref{DualClockAlgebra}), and then the quantum states form the left modules of the representations. Because the algebras (\ref{ClockAlgebra}) and (\ref{DualClockAlgebra}) commute with each other, we can look for representations for them separately, and the complete representation is a tensor product.

\section{Quantization on $\Sigma_1=T^2$}
\label{sec:g1}

\begin{figure}
\centering
\includegraphics[width=0.35\textwidth]{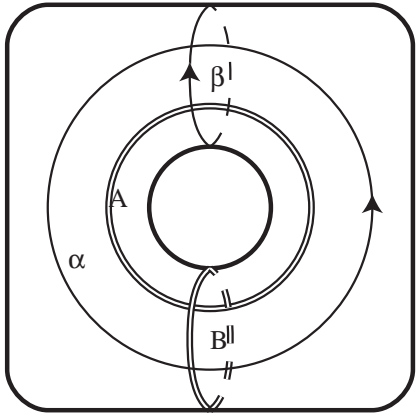}
\caption{Fundamental group generators and MCG generators for $\Sigma_1$. The fundamental group generators are denoted by oriented loops with single thin lines, and labeled by Greek letters; the MCG generators are the unoriented loops with double lines, and labeled by Roman letters.}
\label{fgg1}
\end{figure}

Now we proceed to quantize the theory on a torus. We will need the following topological facts. For a torus, the fundamental group $\pi_1(T^2)$ is abelian and generated by two loops $\bar\alpha,\bar\beta$, with $\bar\alpha\bar\beta=\bar\beta\bar\alpha$. The MCG of the torus, $\mathrm{MCG}(T^2)$ is generated by a pair of Dehn twists, $\A,\B$, which can be presented in Fig.\ref{fgg1} as the loops $A, B$. (In the following, we always pick MCG generators in such a way that they can be presented as simple loops, and for a loop $C$, we denote the corresponding MCG generator by $\C$.) To derive how they operate on loops, we take this convention: after performing a Dehn twist, a loop turns left when it hits the representative loop of the Dehn twist, and goes along the Dehn twist loop, until it comes back to the turning point and continues its original path. With this convention, the MCG generators act on the the fundamental group generators as
\begin{subequations}
\begin{align}
\A(\bar\alpha,\bar\beta)=&(\bar\alpha,\bar\beta\bar\alpha),\\
\B(\bar\alpha,\bar\beta)=&(\bar\beta^{-1}\bar\alpha,\bar\beta).
\end{align}
\label{MCGG1Classical}
\end{subequations}
This group is actually $\mathrm{SL}(2,\mathbb{Z})$, so the relations are
\meg{
&\A\B\A=\B\A\B,\\
&(\B\A\B)^4=1.
}

Let us consider the representation of holonomy group first. This group is generated by $\alpha$ and $\beta$ with $\alpha\beta=\beta\alpha\exp\left(-\frac{\i2\pi}{k}\right)$. In the representation where $\beta$ is diagonal, $\alpha,\beta$ has the block-diagonal form: (From here on, index of every series starts from 0.)
\se{
\beta=\mathrm{diag}\{\tilde\beta(k,\theta_0^\beta),\ldots,\tilde\beta(k,\theta_{r-1}^\beta)\},\quad\alpha=\mathrm{diag}\{\tilde\alpha(k,\theta_0^\alpha),\ldots,\tilde\alpha(k,\theta_{r-1}^\alpha)\},
}
where each block forms irreducible representation and is determined by two parameters
\se{
\tilde\beta(k,\theta^\beta)=\pmx{1\\&\omega\\&&\ddots\\&&&\omega^{p-1}}e^{\i\theta^\beta},\quad
\tilde\alpha(k,\theta^\alpha)=\pmx{&&&1\\1\\&\ddots\\&&1}e^{\i\theta^\alpha},
\label{alphabeta}
}
where $\omega=\exp\left(\frac{\i2\pi}{k}\right)$, $k=p/q$ with $p,q$ coprime, and each block is $p\times p$ dimensional. See appendix A for a proof that this is the most general irreducible representation of the clock algebra. Note that we have not used up all the freedom of unitary transformation on these matrices. While keeping the above form, we can do a cyclic permutation of basis, which changes $\theta^\beta$ by a multiple of $\frac{2\pi}{p}$, and we can shift the phases of the set of basis, which changes $\theta^\alpha$ by a multiple of $\frac{2\pi}{p}$. Thus the phase parameters can be restricted as $\theta^\beta,\theta^\alpha\in[0,\frac{2\pi}{p})$.

Then we need to find representation for the MCG generators, whose operation on the holonomies is derived from their classical operation (\ref{MCGG1Classical}). The relations that $\A$ should satisfy are
\se{
\A^\dag\alpha\A=\alpha,\quad \A^\dag\beta\A=\exp(b+a)=\beta\alpha\omega^{-1/2}.
\label{AQuantumEqn}
}
To solve these equations, we decompose $\A$ into $r\times r$ blocks of $p\times p$ elements as the holonomies. It turns out each block is given by
\se{
\A_{mn}=u_{mn}^\A\tilde\A(k,\theta_n^\alpha,\theta^\A),
}
where $u_{mn}^\A$ is a complex number, and components of $p\times p$ matrix $\tilde\A$ is given by
\se{
\tilde\A(k,\theta_n^\alpha,\theta^\A)_{ij}=\frac{1}{\sqrt p}\omega^{-(i-j)^2/2}e^{\i(i-j)\theta_n^\alpha}e^{\i\theta^\A}.
\label{ABlock}
}
The parameter $\theta^\A$, which does not depend on the block indices $m,n$, is redundant, because it can be absorbed into $u_{mn}^\A$. The reason for this redundancy will be clear later. In addition, (\ref{ABlock}) solves the equations (\ref{AQuantumEqn}) provided it is periodic with respect to the two indices. Changing $i$ from $0$ to $p$ in $\tilde\A_{ij}$ gives the condition $\omega^{-p^2/2}e^{\i p\theta_n^\alpha}=\exp[\i2\pi(-pq/2+p\theta_n^\alpha/2\pi)]=1$. This means if $p$ or $q$ is even, then $\theta_n^\alpha$ are multiples of $\frac{2\pi}{p}$; if $p,q$ are both odd, then $\theta_n^\alpha$ are multiples of $\frac{2\pi}{p}$ plus $\frac{\pi}{p}$. On the other hand, we knew $0\le\theta_n^\alpha<\frac{2\pi}{p}$, so $\theta_n^\alpha$'s are uniquely determined by $k$,
\sen{
\theta_n^\alpha=\frac{\pi}{p}\cdot\Delta, \quad \Delta=\left\{\begin{array}{ll}0, & \textrm{$p$ or $q$ even,}\\
1, & \textrm{$p$ and $q$ odd.}\end{array}\right.
}
In other words, the representations must have the form $\alpha=I\otimes\tilde\alpha(k,\frac{\pi\Delta}{p}),\beta=I\otimes\tilde\beta(k,\frac{\pi\Delta}{p}),\A=U^\A\otimes\tilde\A(k,\frac{\pi\Delta}{p},\theta^\A)$, where $U^\A$ is any unitary matrix. For $\B$, we need
\se{
\B^\dag\alpha\B=\beta^{-1}\alpha\omega^{1/2},\quad \B^\dag\beta \B=\beta.
}
The solution is
\se{
\B_{mn}=u_{mn}^\B\tilde\B(k,\theta_n^\beta,\theta^\B),
}
where
\sen{
\tilde\B(k,\theta_n^\beta,\theta^\B)_{ij}=\omega^{i^2/2}e^{\i i\theta_n^\beta}\delta_{i-j}e^{\i\theta^\B}.
}
By the same periodicity argument, we have
\se{
\theta_n^\beta=\theta_n^\alpha=\frac{\pi}{p}\cdot\Delta, \quad \Delta=\left\{\begin{array}{ll}0, & \textrm{$p$ or $q$ even,}\\
1, & \textrm{$p$ and $q$ odd.}\end{array}\right.
\label{DefinitionDelta}
}
So $\B$ also have the form $\B=U^\B\otimes\tilde\B(k,\frac{\pi\Delta}{p},\theta^\B)$.

Since MCG of $T^2$ is $\mathrm{SL}(2, \mathbb{Z})$, its generators $\A,\B$ should satisfy the relations $\A\B\A=\B\A\B, (\B\A\B)^4=1$. Because of the redundant parameters $\theta^\A$ and $\theta^\B$, we can require that the two parts of direct production satisfy the relations separately. So $U^\A$ and $U^\B$ are generators of an arbitrary unitary representation of the MCG. $\tilde\A$ and $\tilde\B$ are specified by some parameters, and we need to check how the relations constrain those parameters. For the first relation, the right hand side is
\sen{
(\tilde\B\tilde\A\tilde\B)_{ij}=\frac{1}{\sqrt{p}}\omega^{i^2/2+j^2/2-(i-j)^2/2}e^{\i(i+j)\pi\Delta/p+\i(i-j)\pi\Delta/p}e^{\i(2\theta^\B+\theta^\A)}=\frac{1}{\sqrt{p}}\omega^{ij}e^{\i2i\pi\Delta/p}e^{\i(2\theta^\B+\theta^\A)},
}
and the left hand side is
\men{
(\tilde\A\tilde\B\tilde\A)_{ij}=&\frac{1}{p}\sum_{l=0}^{p-1}\omega^{-(i+j-l)^2/2}\omega^{ij}e^{\i(i-j+l)\pi\Delta/p}e^{\i(2\theta^\A+\theta^\B)}\\
=&\frac{1}{p}\sum_{l=0}^{p-1}\omega^{-l^2/2}\omega^{ij}e^{\i(l+2i)\pi\Delta/p}e^{\i(2\theta^\A+\theta^\B)}\\
=&\frac{1}{p}\sum_{l=0}^{p-1}\omega^{-l^2/2+\Delta l/2q}\omega^{ij}e^{\i2i\pi\Delta/p}e^{\i(2\theta^\A+\theta^\B)}.
}
Comparing the two sides, we find a condition for $\theta^\A, \theta^\B$
\se{
e^{\i(\theta^\B-\theta^\A)}=\frac{1}{\sqrt{p}}\sum_{l=0}^{p-1}\exp\left(\frac{\i\pi(-ql^2+\Delta l)}{p}\right).
\label{RelationOne}
}
The right hand side is a quadratic Gauss sum, which is analytically expressed in term of some number-theoretical function. The exact expression will not be presented here. The important fact is, this equation is satisfiable for all possible $p$ and $q$'s ($p,q$ odd, or $p$ even $q$ odd, or $p$ odd $q$ even). One can check that with the help of the term with $\Delta$, the two sides always have the same abstract values, and the phases can be matched by the yet-unknown $\theta_\A$ and $\theta_\B$. For the second relation $(\B\A\B)^4=1$,
\men{
(\tilde\B\tilde\A\tilde\B)_{ij}=&\frac{1}{\sqrt{p}}\omega^{ij}e^{\i2i\pi\Delta/p}e^{\i(2\theta^\B+\theta^\A)},\\
((\tilde\B\tilde\A\tilde\B)^2)_{ij}=&\frac{1}{p}\sum_{l=0}^{p-1}\omega^{il}\omega^{lj}e^{\i(2i+2l)\pi\Delta/p}e^{\i2(2\theta^\B+\theta^\A)}\\
=&\delta_{(q(i+j)+\Delta)\m p}e^{\i2i\pi\Delta/p}e^{\i2(2\theta^\B+\theta^\A)},\\
((\tilde\B\tilde\A\tilde\B)^4)_{ij}=&\delta_{(i-j)}e^{-\i2\pi\Delta q'/p}e^{\i4(2\theta^\B+\theta^\A)},
}
where $q'$ satisfies $qq'\equiv1\mod{p}$. So we get the another condition on $\theta^\A, \theta^\B$
\se{
e^{-\i2\pi\Delta q'/p}e^{\i4(\theta^\A+2\theta^\B)}=1.
\label{RelationTwo}
}
The two conditions (\ref{RelationOne}) and (\ref{RelationTwo}) will determine $\theta^\A$ and $\theta^\B$ up to 12 choices, but different choices are equivalent. Actually, if $\theta^\A$ and $\theta^\B$ are a set of solution, then $\theta^\A+n\frac{2\pi}{12}$ and $\theta^\B+n\frac{2\pi}{12}$ also solve (\ref{RelationOne}) and (\ref{RelationTwo}). The difference between these two set of solutions is nothing but an abelian representation of the MCG, so it can be absorbed into $U^\A$ and $U^\B$.

We still need to find representation for the dual clock algebra (\ref{ClockAlgebra}) for LGTs and representation for the action of MCG to the LGTs. But these are exactly same algebras as those solved above, except $k$ is replaced by $1/k$, or equivalently, $p,q$ are exchanged. Thus irreducible unitary representation for the dual clock algebra is $q$-dimensional, phase parameters $\theta_n^\rho$ and $\theta_n^\sigma$ are uniquely determined to be $\frac{\pi}{q}\cdot\Delta$, and counterparts of (\ref{RelationOne}) and (\ref{RelationTwo}) give some solution to the phase parameters $\hat\theta^\A$ and $\hat\theta^\B$. The full representation for the algebras is
\men{
&\alpha=I_r\otimes\tilde\alpha\left(k,\pi\Delta/p\right)\otimes I_q,\quad\beta=I_r\otimes\tilde\beta\left(k,\pi\Delta/p\right)\otimes I_q,\\
&\rho=I_r\otimes I_p\otimes\tilde\alpha\left(1/k,\pi\Delta/q\right),\quad\sigma=I_r\otimes I_p\otimes\tilde\beta\left(1/k,\pi\Delta/q\right)\\
&\A=U^\A\otimes\tilde\A\left(k,\pi\Delta/p,\theta^\A\right)\otimes\tilde\A\left(1/k,\pi\Delta/q,\hat\theta^\A\right),\quad\B=U^\B\otimes\tilde\B\left(k,\pi\Delta/p,\theta^\B\right)\otimes\tilde\B\left(1/k,\pi\Delta/q,\hat\theta^\B\right).
}

Unlike Ref.\cite{Bos:1989wa}, value of $k$ is not restricted after imposing MCG. However, representing MCG still affect the representation of the holonomy group and the LGT group, because in this process, the phase parameters $\theta_n^\alpha$ and $\theta_n^\beta$ are determined as in (\ref{DefinitionDelta}), which are otherwise free to change within $[0,\frac{2\pi}{p})$, and $\theta_n^\rho, \theta_n^\sigma$ are restricted similarly.

\section{Quantization on $\Sigma_2$}

\begin{figure}
\centering
\includegraphics[width=0.65\textwidth]{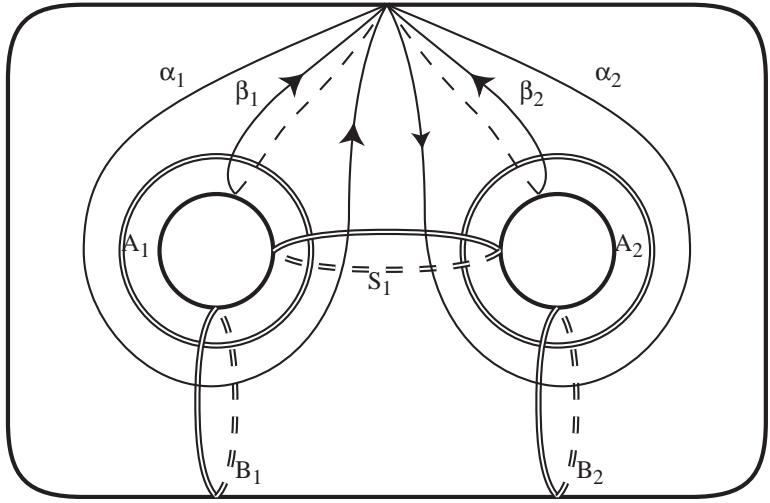}
\caption{Fundamental group generators and MCG generators for $\Sigma_2$. The fundamental group generators are denoted by oriented loops with single thin lines, and labeled by Greek letters; the MCG generators are the unoriented loops with double lines, and labeled by Roman letters.}
\label{fgg2}
\end{figure}

The relevant topological properties of $\Sigma_2$ are the following. Fundamental group of $\Sigma_2$ has 4 generators $\bar\alpha_1,\bar\beta_1,\bar
\alpha_2,\bar\beta_2$ with one relation $\bar\alpha_1^{-1}\bar\beta_1\bar\alpha_1\bar\beta_1^{-1}\bar\alpha_2^{-1}\bar\beta_2\bar\alpha_2\bar\beta_2^{-1}=1$. The MCG of $\Sigma_2$, $\mathrm{MCG}(\Sigma_2)$, is generated by five Dehn twists, $\A_1,\B_1,\A_2,\B_2,\S$, as are drawn in Fig.\ref{fgg2}. Their operation on the loops are derived to be
\begin{subequations}
\begin{align}
\A_1(\bar\alpha_1,\bar\beta_1,\bar\alpha_2,\bar\beta_2)&=(\bar\alpha_1,\bar\beta_1\bar\alpha_1,\bar\alpha_2,\bar\beta_2),\label{op1}\\
\B_1(\bar\alpha_1,\bar\beta_1,\bar\alpha_2,\bar\beta_2)&=(\bar\beta_1^{-1}\bar\alpha_1,\bar\beta_1,\bar\alpha_2,\bar\beta_2),\\
\A_2(\bar\alpha_1,\bar\beta_1,\bar\alpha_2,\bar\beta_2)&=(\bar\alpha_1,\bar\beta_1,\bar\alpha_2,\bar\beta_2\bar\alpha_2),\\
\B_2(\bar\alpha_1,\bar\beta_1,\bar\alpha_2,\bar\beta_2)&=(\bar\alpha_1,\bar\beta_1,\bar\beta_2\bar\alpha_2^{-1},\bar\beta_2),\\
\S(\bar\alpha_1,\bar\beta_1,\bar\alpha_2,\bar\beta_2)&=(\bar\alpha_1\bar\beta_1^{-1}\bar\beta_2,\bar\beta_1,\bar\beta_2^{-1}\bar\beta_1\bar\alpha_2,\bar\beta_2),\label{op5}
\end{align}
\end{subequations}
and they satisfy the following relations \cite{Birman:1971}
\begin{subequations}
\begin{align}
&[\A_1,\A_2]=[\A_1,\B_2]=[\B_1,\A_2]=[\B_1,\B_2]=[\B_1,\S]=[\B_2,\S]=1,\label{relation1}\\
&\A_1\B_1\A_1=\B_1\A_1\B_1,\quad \A_2\B_2\A_2=\B_2\A_2\B_2,\quad \A_1\S\A_1=\S\A_1\S,\quad \A_2\S\A_2=\S\A_2\S,\label{relation2}\\
&(\B_1\A_1\S)^4=\B_2^2,\label{relation3}\\
&[\B_2\A_2\S\A_1\B_1\B_1\A_1\S\A_2\B_2,\B_1]=1,\label{relation4}\\
&(\B_2\A_2\S\A_1\B_1\B_1\A_1\S\A_2\B_2)^2=1.\label{relation5}
\end{align}
\end{subequations}

By the convention of Fig.\ref{fgg2}, with $a_i=\int_{\alpha_i}A, b_i=\int_{\beta_i}A$, the Chern-Simons action is
\sen{
I_{\textrm{CS}}=\frac{k}{2\pi}\int dt(a_1\partial_tb_1+a_2\partial_tb_2)
}
which results in the commutators
\sen{
[a_1,b_1]=[a_2,b_2]=\frac{-\i2\pi}{k}
}
and any other pair commutes.

Like the previous example, let us only consider representation of the holonomy group for now. From the above commutation relations, the holonomy group generators satisfy the clock algebra,
\se{
\alpha_1\beta_1=\beta_1\alpha_1\omega^{-1}, \quad \alpha_2\beta_2=\beta_2\alpha_2\omega^{-1},
}
and any other pair commutes. Same as in the previous section, we have the following equations for the MCG generators,
\sen{
\B_1^\dag\alpha_1\B_1=\beta_1^{-1}\alpha_1\omega^{1/2},\quad \B_2^\dag\alpha_2\B_2=\beta_2\alpha_2\omega^{1/2},\quad \A_1^\dag\beta_1\A_1=\beta_1\alpha_1\omega^{-1/2},\quad \A_2^\dag\beta_2\A_2=\beta_2\alpha_2^{-1}\omega^{-1/2},
}
and they commute with the rest of the holonomy group generators.

The equations we have listed so far are just two copies of their counterparts in section \ref{sec:g1}, so the solution of the generators is just a direct product of the solutions in section \ref{sec:g1},
\me{
&\beta_1=I_r\otimes\tilde\beta_1,\quad\beta_2=I_r\otimes\tilde\beta_2,\quad \alpha_1=I_r\otimes\tilde\alpha_1,\quad\alpha_2=I_r\otimes\tilde\alpha_2,\\
&\B_1=U_1^\B\otimes\tilde\B_1,\quad\B_2=U_2^\B\otimes \tilde\B_2,\quad \A_1=U_1^\A\otimes\tilde\A_1,\quad\A_2=U_2^\A\otimes \tilde\A_2,
}
where
\men{
&\tilde\beta_1=\tilde\beta\left(k,\pi\Delta/p\right)\otimes I_p,\quad \tilde\beta_2=I_p\otimes\tilde\beta\left(k,\pi\Delta/p\right),\\ &\tilde\alpha_1=\tilde\alpha\left(k,\pi\Delta/p\right)\otimes I_p,\quad \tilde\alpha_2=I_p\otimes\tilde\alpha\left(k,\pi\Delta/p\right),\\
&\tilde\B_1=\tilde\B\left(k,\pi\Delta/p,\theta^\B_1\right)\otimes I_p,\quad \tilde\B_2=I_p\otimes\tilde\B\left(k,\pi\Delta/p,\theta^\B_2\right),\\ &\tilde\A_1=\tilde\A\left(k,\pi\Delta/p,\theta^\A_1\right)\otimes I_p,\quad \tilde\A_2=I_p\otimes\tilde\A\left(k,\pi\Delta/p,\theta^\A_2\right).
}
The new generator $\S$ are determined by the equations
\sen{
\S^\dag\alpha_1\S=\alpha_1\beta_1^{-1}\beta_2\omega^{-1/2},\quad \S^\dag\alpha_2\S=\beta_2^{-1}\beta_1\alpha_2\omega^{1/2},\quad \S^\dag\beta_1\S=\beta_1,\quad \S^\dag\beta_2\S=\beta_2
}
and the solution is
\sen{
\S=U^\S\otimes\tilde\S(k,\theta^\S),
}
where
\sen{
\tilde\S(k,\theta^\S)_{i_1j_1,i_2j_2}=\omega^{(i_2-i_1)^2/2}\delta_{i_1-j_1}\delta_{i_2-j_2}e^{\i\theta^\S}.
}
This matrix is periodic only if one of $p,q$ are even, that is, $\Delta=0$. This is a non-trivial ``quantization condition'' for $k$ on $\Sigma_2$. One can see that if we restrict $k$ to be integer-valued, then this condition agree with that in \cite{Bos:1989wa}, which is $k$ must be an even integer.

Now we check the relations (\ref{relation1})-(\ref{relation5}). (\ref{relation1}) is automatically satisfied. (\ref{relation2}) gives the same equation as (\ref{RelationOne}) except now $\Delta=0$,
\se{
e^{\i(\theta^\B_1-\theta^\A_1)}=e^{\i(\theta^\B_2-\theta^\A_2)}=e^{\i(\theta^\S-\theta^\A_1)}=e^{\i(\theta^\S-\theta^\A_2)}=\frac{1}{\sqrt{p}}\sum_{l=0}^{p-1}\exp\left(\frac{-\i\pi ql^2}{p}\right)\label{result1}
}
which means $\theta^\A_1=\theta^\A_2, \theta^\B_1=\theta^\B_2=\theta^\S$, and these two angles are related by this equation. So there is really only one free angle parameter left. Consider the relation (\ref{relation3}),
\men{
((\tilde\B_1\tilde\A_1\tilde\S)^2)_{i_1j_1,i_2j_2}=&\delta_{i_1+j_1-j_2}\delta_{i_2-j_2}\omega^{-j_1j_2+j_2^2}e^{\i2(\theta^\B_1+\theta^\A_1+\theta^\S)}\\
((\tilde\B_1\tilde\A_1\tilde\S)^4)_{i_1j_1,i_2j_2}=&\delta_{i_1-j_1}\delta_{i_2-j_2}\omega^{j_2^2}e^{\i4(\theta^\B_1+\theta^\A_1+\theta^\S)}
}
so (\ref{relation3}) gives
\se{
e^{\i4(\theta^\B_1+\theta^\A_1+\theta^\S)-\i2\theta^\B_2}=1\label{result2}
}
and this will fix all the angles up to 10 choices. Next consider the relation (\ref{relation4}),
\men{
(\tilde\B_2\tilde\A_2\tilde\S\tilde\A_1\tilde\B_1)_{i_1j_1,i_2j_2}=&\frac{1}{p}\omega^{i_1j_1-i_1j_2+i_2j_2}e^{\i(\theta^\A_1+\theta^\A_2+\theta^\B_1+\theta^\B_2+\theta^\S)}\\
(\tilde\B_2\tilde\A_2\tilde\S\tilde\A_1\tilde\B_1\tilde\B_1\tilde\A_1\tilde\S\tilde\A_2\tilde\B_2)_{i_1j_1,i_2j_2}=&\delta_{i_1+j_1}\delta_{i_2+j_2}e^{\i2(\theta^\A_1+\theta^\A_2+\theta^\B_1+\theta^\B_2+\theta^\S)}
}
It can be checked now that this matrix commutes with $\tilde\B_1$. Multiplying $\tilde\B_1$ from left or from right just adds the factor $\omega^{i_1^2/2}$ or $\omega^{j_1^2/2}$ respectively, and they are the same because of the first delta function. So the relation (\ref{relation4}) poses no condition on the angles. The last relation (\ref{relation5}) is
\sen{
((\tilde\B_2\tilde\A_2\tilde\S\tilde\A_1\tilde\B_1\tilde\B_1\tilde\A_1\tilde\S\tilde\A_2\tilde\B_2)^2)_{i_1j_1,i_2j_2}=\delta_{i_1-j_1}\delta_{i_2-j_2}e^{\i4(\theta^\A_1+\theta^\A_2+\theta^\B_1+\theta^\B_2+\theta^\S)}
}
so
\se{
e^{\i4(\theta^\A_1+\theta^\A_2+\theta^\B_1+\theta^\B_2+\theta^\S)}=1\label{result3}
}
Given (\ref{result1}) and (\ref{result2}), this condition is redundant. The nontrivial conditions are (\ref{result1}) and (\ref{result2}), and they have $10$ distinct solutions, which are equivalent because the difference between the solutions are merely an abelian representation of MCG.

By repeating the above calculation with $k$ replaced by $1/k$, we can get the representation for the dual algebra satisfied by LGT generators. In particular, the quantization condition for $k$, which is one of $p,q$ must be even, is symmetric with respect to $p$ and $q$, so the dual algebra will not give additional restriction on $k$.

\section{Quantization on $\Sigma_g$, $g\ge3$}

\begin{figure}
\centering
\includegraphics[width=1.0\textwidth]{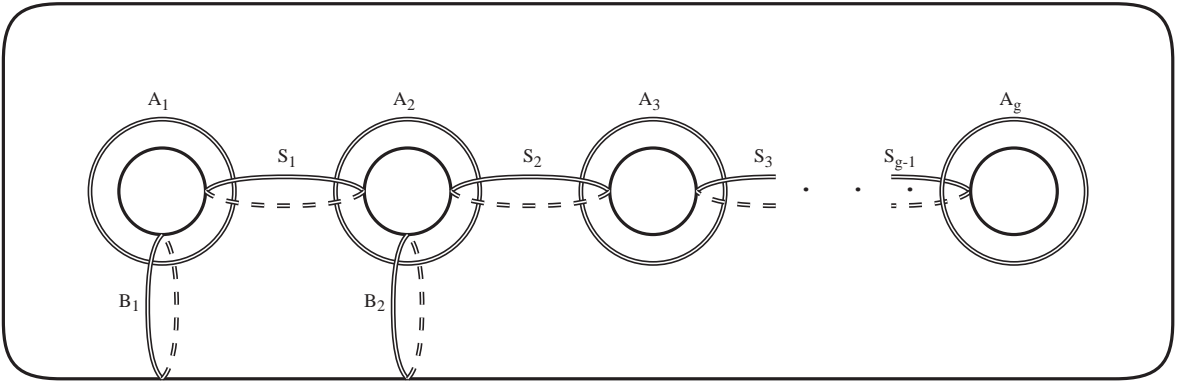}
\caption{MCG generators for $\Sigma_g, g\ge3$.}
\label{fggg}
\end{figure}

Fundamental group and MCG of $\Sigma_g$ with $g\ge3$ have the following presentation. The fundamental group $\pi_1(\Sigma_g)$ is generated by $2g$ loops $\bar\alpha_1,\ldots,\bar\alpha_g,\bar\beta_1,\ldots,\bar\beta_g$, with one relation $\bar\alpha_1^{-1}\bar\beta_1\bar\alpha_1\bar\beta_1^{-1}\cdots\bar\alpha_g^{-1}\bar\beta_g\bar\alpha_g\bar\beta_g^{-1}=1$. For $g\ge3$, MCG have $2g+2$ generators, which are $\A_n,n=1,\ldots,g$, $\S_n,n=1,\ldots,g$, and $\B_n,n=1,2$, as shown in Fig.\ref{fggg}. An explicit presentation of MCG is given in \cite{Wajnryb:1983}. It takes the following form in our convention
\meg{
&[\C,\C']=1,\quad\textrm{when $\#(C,C')=0$},\label{g3r1}\\
&\C\C'\C=\C'\C\C',\quad\textrm{when $\#(C,C')=\pm1$},\quad&\textrm{(braid relation)},\label{g3r2}\\
&(\S_1\A_1\B_1)^4=\E_0\B_2,\quad&\textrm{(3-chain relation)},\label{g3r3}\\
&\E_2\E_1\B_2=\E_3\S_2\S_1\B_1,\quad&\quad\textrm{(lantern relation)},\label{g3r4}\\
&[\A_g\S_{g-1}\A_{g-1}\cdots \S_1\A_1\B_1\B_1\A_1\S_1\cdots \A_{g-1}\S_{g-1}\A_g,\B_g]=1,\quad&\textrm{(hyperelliptic relation)},\label{g3r5}
}
where
\men{
\E_0=&(\A_2\S_1\A_1\B_1\B_1\A_1\S_1\A_2)\B_2(\A_2\S_1\A_1\B_1\B_1\A_1\S_1\A_2)^{-1},\\
\E_1=&(\A_2\S_2\S_1\A_2)^{-1}\B_2(\A_2\S_2\S_1\A_2),\\
\E_2=&(\A_1\S_1\B_1\A_1)^{-1}\E_1(\A_1\S_1\B_1\A_1),\\
\E_3=&(\A_3\S_2\A_2\S_1\A_1\U\B_1^{-1}\A_1^{-1}\S_1^{-1}\A_2^{-1})\B_2(\A_3\S_2\A_2\S_1\A_1\U\B_1^{-1}\A_1^{-1}\S_1^{-1}\A_2^{-1})^{-1},\\
\U=&(\A_3\S_2)^{-1}\E_1^{-1}(\A_3\S_2),
}
and $\B_{n+2}$ is computed from $\B_n, \B_{n+1}$ and the generators by induction
\se{
\B_{n+2}=\W_n\B_n\W_n^{-1},\label{induction}
}
where
\sen{
\W_n=(\A_n\S_n\A_{n+1}\B_{n+1})(\S_{n+1}\A_{n+2}\A_{n+1}\S_{n+1})(\S_n\A_{n+1}\A_n\S_n)(\B_{n+1}\A_{n+1}\S_{n+1}\A_{n+2}).
}

The holonomies are $g$ copies of quantum torus, and they satisfy clock algebra in pairs. Representation of the $2g+2$ MCG generators can also be derived from the their actions on the holonomies
\men{
&\beta_n=I_r\otimes\tilde\beta_n,\quad \alpha_n=I_r\otimes\tilde\alpha_n,\\
&\A_n=U^\A_n\otimes\tilde\A_n,\quad \S_n=U^\S_n\otimes\tilde\S_n,\quad \B_n=U^\B_n\otimes\tilde\B_n,
}
where
\meg{
&\tilde\beta_n=I_p^{\otimes n-1}\otimes\tilde\beta(k,0)\otimes I_p^{\otimes g-n},\\
&\tilde\alpha_n=I_p^{\otimes n-1}\otimes\tilde\alpha(k,0)\otimes I_p^{\otimes g-n},\\
&\tilde\A_n=I_p^{\otimes n-1}\otimes\tilde\A(k,0,\theta^\A_n)\otimes I_p^{\otimes g-n},\\
&\tilde\S_n=I_p^{\otimes n-1}\otimes\tilde\S(k,0,\theta^\S_n)\otimes I_p^{\otimes g-n-1},\\
&\tilde\B_n=I_p^{\otimes n-1}\otimes\tilde\B(k,0,\theta^\B_n)\otimes I_p^{\otimes g-n}.
\label{Bn}
}
In this process, we get the same quantization condition on $k$, which is one of $p,q$ must be even.

We still need to make sure the relations (\ref{g3r1})-(\ref{g3r5}) are satisfied. (\ref{g3r1}) is trivially satisfied. Braid relations (\ref{g3r2}) give as before
\se{
\theta^\A_n=\theta^\A,\ \theta^\B_n=\theta^\S_n=\theta^\B,\ e^{\i(\theta^\B-\theta^\A)}=\frac{1}{\sqrt{p}}\sum_{l=0}^{p-1}\exp\left(\frac{-\i\pi ql^2}{p}\right)
}
There leaves only one undetermined phase in the MCG generators. The 3-chain relation (\ref{g3r3}) is
\sen{
(\S_1\A_1\B_1)^4=(\A_2\S_1\A_1\B_1\B_1\A_1\S_1\A_2)\B_2(\A_2\S_1\A_1\B_1\B_1\A_1\S_1\A_2)^{-1}\B_2
}
Note that every element in this relation have the same form as in the previous section, so the results can be used. From (\ref{relation4}), we know $[\A_2\S_1\A_1\B_1\B_1\A_1\S_1\A_2,\B_2]=1$, by conjugation and commutation, this relation reduce to the previously proven relation (\ref{relation3}), $(\B_1\A_1\S_1)^4=\B_2^2$. About the lantern relation (\ref{g3r4})
\sen{
\E_2\E_1\B_2=\E_3\S_2\S_1\B_1
}
After some calculation, we find
\men{
(\tilde\E_2\tilde\E_1\tilde\B_2)_{i_1j_1,\ldots,i_gj_g}=&\omega^{j_1^2+j_2^2+j_3^2-j_1j_2-j_2j_3}e^{\i3\theta^\B}\delta_{i_1-j_1}\delta_{i_2-j_2}\delta_{i_3-j_3}\cdots\\
(\tilde\E_3\tilde\S_2\tilde\S_1\tilde\B_1)_{i_1j_1,\ldots,i_gj_g}=&\omega^{j_1^2+j_2^2+j_3^2-j_1j_2-j_2j_3}e^{\i4\theta^\B}\delta_{i_1-j_1}\delta_{i_2-j_2}\delta_{i_3-j_3}\cdots
}
so $\theta^\B=0$. This fix the remaining freedom in the MCG generators. To check the hyperelliptic relation (\ref{g3r5}),
\sen{
[\A_g\S_{g-1}\A_{g-1}\cdots \S_1\A_1\B_1\B_1\A_1\S_1\cdots \A_{g-1}\S_{g-1}\A_g,\B_g]=1
}
by (\ref{induction}), we find $\tilde\B_n$ takes the form of (\ref{Bn}) for all $n$, and
\sen{
(\tilde\A_g\tilde\S_{g-1}\tilde\A_{g-1}\cdots\tilde\S_1\tilde\A_1\tilde\B_1\tilde\B_1\tilde\A_1\tilde\S_1\cdots\tilde\A_{g-1}\tilde\S_{g-1}\tilde\A_g)_{i_1j_1,\ldots,i_gj_g}=\delta_{i_1+j_1}\cdots\delta_{i_g+j_g}e^{\i g(\theta^\A+\theta^\B)}
}
Obviously this commutes with $\tilde\B_g$.

Like in the previous section, representation of the dual clock algebra give no additional quantization condition on $k$.

\section{Discussion}

To summarize the result, we find that the $\mathrm{U}(1)$ Chern-Simons theory defined on $\mathbb{R}\times\Sigma_g$ is quantizable, if we require the quantum states to form representations for the deformed holonomy group, the deformed LGT group, and the MCG. Explicit, finite dimensional representation of these groups are found. The parameter $k$ is quantized in an interesting manner. For $\Sigma_1$, $k$ can take any nonzero rational value, while for $\Sigma_g$ with $g\ge2$, $k$ must be a rational number with either its numerator or denominator being even. The representations are unique in the sense that, apart from a choice of arbitrary unitary representation of MCG, the representations of the discrete groups (holonomy group + LGT group + MCG) are completely fixed.

Uniqueness of the representation of these discrete groups is an interesting result. In general, when one considers gravity with some non-trivial space-time topology, it is expected that some theta angle parameters, or more complicated non-abelian parameters, will arise. But in our toy model, which can be regarded as three dimensional Chern-Simons gravity with gauge group $\mathrm{U}(1)$ instead of some non-abelian gauge group \cite{Witten:1988hc}, although some theta angle parameters appear in representation of the clock algebra, they disappear after representing the MCG.

The representations of the discrete groups on $\Sigma_g$ are found to be $r(pq)^g$ dimensional, where $r$ is the dimensionality of an arbitrary unitary representation of the MCG, and $k=p/q$ with $p,q$ coprime. This can be compared with the result of the path-integral quantization in Ref.\cite{Bos:1989wa}. If we take $r=1$ and $k$ integer-valued, it is easy to check that, the $k^g$ dimensional dimensional representation of the holonomy group from this work is exactly the same representation formed by states in the $k^g$ dimensional Hilbert space found in Ref.\cite{Bos:1989wa}(see Eqn. (23) and (24) therein). As a consequence, the representations of MCG in this work and that in Ref.\cite{Bos:1989wa} are also the same.

The $r(pq)^g$ dimensional Hilbert space that we find can be viewed as a direct product of $g$ $(pq)$-dimensional subspaces, and one $r$-dimensional subspace. Each $(pq)$-dimensional subspace is associated with one specific handle on $\Sigma_g$, and the $r$-dimensional subspace forms an extra representation of the MCG. Due to this decomposed structure, we can consider pinching of the handles, in which some handles are shrank to marked points. Because the quantization condition on $k$ is stronger for $g\ge2$ than for $g=1$, an allowed $k$ value on a higher genus surface will not pose any problem on a lower genus surface. Without loss of generality, assume the first handle is pinched. The holonomy around the remaining marked points is $\alpha_1\beta_1\alpha_1^{-1}\beta_1^{-1}$, which is $\exp\left(-\frac{\i2\pi}{k}\right)$ by Eqn.(\ref{DualClockAlgebra}). This means in a quantum state, after the pinching, all information about holonomies of the first handle is lost. The same is true for information about LGTs, with the same reason. Thus the $(pq)^g$-dimensional subspace becomes $(pq)^{g-1}$ dimensional after pinching. For the $r$-dimensional subspace, the situation is more complicated. In general, representations of MCG of higher genus surface does not reduce to representations of MCG of lower genus surface with marked points, so if pinching is allowed, there may be some extra conditions on the $r$-dimensional representation of MCG.

As shown in previous sections, to impose the large gauge symmetry and large diffeomorphisms, it is impossible to simply reduce the original classical phase space to some invariant subspace of these symmetries. However, it is straightforward, at least classically, to find the invariant phase space under one of the two symmetries. In fact, we chose to represent the large gauge symmetry first, and as a result the $b_1(\Sigma_g)$-dimensional quantum plane, where $b_1$ is the first Betti number, reduces to a $b_1(\Sigma_g)$-dimensional torus times a $\mathbb{Z}^{b_1(\Sigma_g)}$ lattice, both of which are deformed by the canonical commutator to non-commutative spaces. Wave functions take values on these two parts of phase space separately. Then the implementation of MCG gives some non-trivial and rather technical restrictions on the parameters of the theory, as were listed above.

In principle, the other way around, i.e., to impose MCG first should be equally practicable. Indeed, the invariant subspace is the moduli space of Teichm\"uller space of $\Sigma_g$, and wave functions are sections based upon this moduli space. But it is not clear how the wave functions can carry a representation of the large gauge transformations.

\appendix
\section{Irreducible Representations of Clock Algebra}

If $\tilde\beta$ and $\tilde\alpha$ form a representation of the clock algebra, and $\tilde\beta$ is diagonal, their components satisfy
\sen{
\tilde\alpha_{ij}\tilde\beta_j=\tilde\beta_i\tilde\alpha_{ij}\omega^{-1}.
}
So if $\tilde\alpha_{ij}$ is nonzero, $\tilde\beta_i=\omega\tilde\beta_j$. A general solution can be written as
\sen{
\tilde\beta=\pmx{
I_{d_0}\\
&\omega I_{d_1}\\
&&\ddots\\
&&&\omega^{p-1}I_{d_{p-1}}\\
}e^{\i\theta^\beta},\quad
\tilde\alpha=\pmx{
&&&\hat\alpha_{0,p-1}\\
\hat\alpha_{10}\\
&\ddots\\
&&\hat\alpha_{p-1,p-2}
}e^{\i\theta^\alpha},
}
where $d_i$ is the size of each block, and $\hat\alpha_{i,i-1}$ is $d_i\times d_{i-1}$ dimensional. But $\tilde\alpha$ can be unitary only if $d_0=d_1=\cdots=d_{p-1}=d$. While fixing $\tilde\beta$, the allowed unitary transformations have the form $U=\mathrm{diag}\{U_0,\ldots,U_{p-1}\}$, and
\sen{
U^{-1}\tilde\alpha U=\pmx{
&&&U_{0}^{-1}\hat\alpha_{0,p-1}U_{p-1}\\
U_1^{-1}\hat\alpha_{10}U_0\\
&\ddots\\
&&U_{p-1}^{-1}\hat\alpha_{p-1,p-2}U_{p-2}
}e^{\i\theta^\alpha}.
}
We can take all the $U_i$'s to be the same, but perform such unitary transformation $p$ times. One of the $\hat\alpha_{i,i-1}$'s can be diagonalized each time. Finally we get a reducible representation unless $d=1$.

\section*{Acknowledgments}

I am deeply indebted to Gordon Semenoff, Donald Witt and Kristin Schleich for motivating this project, as well as for a sequel of discussion that clarified many aspects of the problem. This work is supported by NSERC of Canada and 4YF of University of British Columbia.

\end{document}